# Classification of step bunching phenomena

Vesselin Tonchev,

Institute of Physical Chemistry, Bulgarian Academy of Sciences,

1113 Sofia, Bulgaria

**Abstract.** The classification of bunching of straight steps on vicinal crystal surfaces identifies two types according to the behavior of the minimal step-step distance in the bunch $l_{min}$ with increasing the number of steps $N$ in it. In the B1-type $l_{min}$ remains constant while in the B2-type it decreases. Both types are illustrated by new results for well-known models. The precise numerical analysis is aimed at the intermediate asymptotic regime where self-similar spatiotemporal patterns develop. In the model of Tersoff et al. the regular step train is destabilized by step-step attraction of infinite range. It is shown that this model belongs to the B1-type and the same time-scaling exponent of 1/5 for $N$, terrace width and bunch width is obtained. An extended set of scaling exponents is obtained from the model of S.Stoyanov of diffusion-limited evaporation affected by electromigration of the adatoms. This model is of B2-type and shows a systematic shift of the exponents with respect to the predictions of the hypothesis for universality classes in bunching thus requiring further modification of it.

**Retrospective.** The most intensive studies of the bunching of straight steps on vicinal crystal surfaces were initiated by the discovery of Latyshev et al. [1] – equally spaced steps gather in groups in result of resistive DC-heating of Si(111)-vicinals (surfaces of a monocrystal that are slightly deviated from the (111)-plane and the result is sequence of straight equidistant monoatomic steps). A theoretical explanation of the phenomenon was suggested by Stoyanov[2] who identified as source of the instability the electromigration of Si-adatoms which causes bias of the surface diffusion and thus uneven contribution of the two terraces adjacent to a step to its motion. In these initial years the theoretical efforts were focused on understanding and predicting the initial stages of the process – the way the instability arises and grows further. More recently were addressed [3-5] also the late stages of the process when the instability is well developed and enters into the so called intermediate asymptotic regime [6] in which the surface morphology becomes self-similar both in space and in time. The adequate description of this regime is given in terms of scaling laws which include a combination of model parameters, length-scale(s) of the phenomenon and time. There are several reasons for the continuing interest in bunching studies: (i) the so called step flow growth mode is the important one from technological point of view; (ii) bunched surfaces are nowadays used as templates for bottom-up strategies to grow nanostructures [7]; (iii) this is a clear and very rich case of surface self-organization after the system is driven out of equilibrium [8] and opposite tendencies compete – surface destabilization due to various kinetic or thermodynamic factors opposed by the omnipresent step-step repulsion which favors the equidistant step distribution.

**Classification.** The classification was introduced recently by Staneva et al. [9]. The step bunching phenomena are classified according to the behavior of the minimal step-step distance in the bunch $l_{min}$ with increasing the number of step $N$ in it as shown in the table below.

| When increasing the number of steps $N$ in the bunch .. | .. the minimal step-step distance in the bunch $l_{min}$ .. | Number of characteristic length-scales |
|---|---|---|
| B1-type | .. remains constant | 1 |
| B2-type | .. decreases | 2 |

There are several reasons that hindered formulation of the classification in the years: (i) the communities that carry out active research on the two types practically do not overlap; (ii) the property of constant $l_{min}$ was not explicitly recognized in the model studies of the B1-type [10,11]; (iii) the number of necessary length-scales to describe the step bunching thoroughly was not instituted firmly in the protocol for the B2-type studies. In what follows I describe briefly the computational protocol and then present the two types with numerical results from well-known models stressing on the new findings.

*Numerical procedure.* – There are two general ways to obtain the system of ordinary differential equations (ODE's) for the velocities of steps. The usual one, of extended Burton-Cabrerra-Frank type (essentially 1D approach), is to deduce these by rigorous considerations - solving the proper diffusion equation on a single terrace with diffusion bias entering the equation, i.e. the drift of adatoms due to electromigration, and/or the asymmetry in kinetics of attachment/detachment entering the boundary conditions (BC) on the steps through unequal kinetic coefficients. The step-step repulsion also enters the BC modifying the equilibrium (reference) concentrations used to calculate the actual deviation from equilibrium. Step velocity is proportional to the diffusive fluxes entering the step from both terraces. Another approach is to construct velocity equation(s) *ad hoc* [12]. Once a system of ODE's is defined it is solved numerically by a suitable routine, usually fourth order Runge-Kutta but other modern integration strategies could be adopted as well [13]. The non-trivial part of the study is to design and implement a procedure to recognize the evolving surface pattern and to extract the information needed [14,15]. Usually not less than 1000 steps (=equations) are included in the calculation in order to ensure smoothness of the quantities that describe the evolution of the system. Our computational protocol [14,15] consists of gathering statistics based on the step-step distances with two monitoring schemes (MS) running simultaneously with the important definition in the background of what is *bunch distance*. We define a step-step distance to be a *bunch distance* always when it is smaller than the initial (vicinal) one, usually denoted by *l*. Other choices are also possible[16] but these are not justified by physical considerations. The first monitoring scheme, MS – I, is designed to follow the temporal evolution of the system. It calculates at every time step of the integration the number of bunches in the system and then the average number of steps in bunch *N*, the average bunch width $L_b$, average terrace width between bunches TW, etc. plus an individual quantity – the global minimal step-step distance in the system $l_{min\_g}$. These quantities are written in files versus time. The second monitoring scheme, MS - II, cumulates separately information for any bunch size - average bunch width $L_b$, average minimal step-step distance in the bunch $l_{min}$, first and last step-step distances $l_1$ and $l_{last}$, see Figures 1 and 6 for some definitions, that would appear during the whole simulation and at every time step this information is updated and written in files versus the bunch size. Thus at every time step integral information is available from MS - II that reflects the whole surface evolution up to this moment. The coincidence of the results from the two schemes for a matching dependence as the bunch width vs. bunch size is considered non-trivial and mutual validation of the two schemes, see Figure 7. Mostly used are the *N* vs. time dependence from MS – I (plus bunch width $L_b$ vs. time when studying B2-type) and $l_{min}$ vs. *N* from MS – II. The combination of the two schemes could be used also for structuring the data from experiments in order to plot it in the same coordinates as the data from calculations and make a direct comparison.

*B1-type.* – Examples for experimental systems that show B1-type step bunching are the Si(113) surface, high temperature annealing of TaC(910) [17], vicinal Ag(111) in electrolyte [18], etc. I will illustrate this type with results from the model of step bunching due a step-step attraction of infinite range, proposed by Tersoff et al.[10], we will call this TE-model. The other limiting case of zero-range step-step attraction called 'sticky steps' was introduced quite recently [19, 20] and the surface slope behavior in this model is still under investigation. The physical origin of the step-step attraction in the TE-model is identified as the strain cumulated during heteroepitaxial growth which remains uncompensated at the steps and forms force monopoles. As a result instability develops mediated by mass diffusion which breaks the equidistant step distribution, thus being a vicinal analogue to the Asaro-Tiller-Grinfeld instability [8]. The equations for step velocity are published[10] and here only a comment is provided instead – this model and two more [12,21] have a specific property of the equations, namely, they consist of two mathematically identical terms with opposite effect, destabilizing and stabilizing, and with different notation of the parameters. The second term introduces the effect of the omnipresent step-step repulsion and the first one – the effect of the emerging step-step attraction. In the original model of Tersoff et al. [10] the energy of step-step repulsion decays inversely proportional to the square of the step-step distance and the energy of step-step attraction increases with the logarithm of the step-step distance – the bigger is the distance between the steps, the larger is the uncompensated strain. The results are obtained with same values of the parameters: $\alpha_1$ = 10 (destabilizing), $\alpha_2$ = 1 (stabilizing), the first one contains the magnitude of the step-step attraction and the second one – the magnitude of step-step repulsion. In Figure 1 are plotted the step trajectories and some of the elements of the monitoring schemes are marked. Figure 2 contains data on the surface profile for well-developed instability – as seen the slope is constant along the surface while on Figures 3 and 4 it is seen that the slope does not change with the increase of the bunch size $N$ and in time. Figure 4 demonstrates that only one time-scaling exponent is found for three different quantities – $N$, terrace width and bunch width and it is 1/5, different from 1/4 as found by Tersoff et al. [10]. The reasons for this difference are still unclear and subject to further studies.

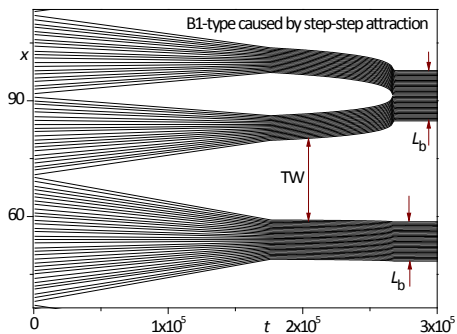

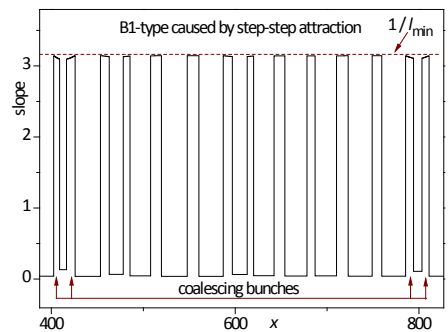

Figure 1. TE-model, step trajectories as obtained when solving numerically the equations for step velocity [10]. It is seen that when two bunches coalesce the resulting bunch has a width being the sum of the bunch widths before the coalescence.

Figure 2. TE-model, surface profile (inverse of the step-step distances) for well-developed instability. The width of the 'peaks' is the bunch width $L_b$. Interesting dynamic phenomenon is observed – when two bunches coalesce the slope of each is lower from the side of the other bunch (the double arrows from the bottom).

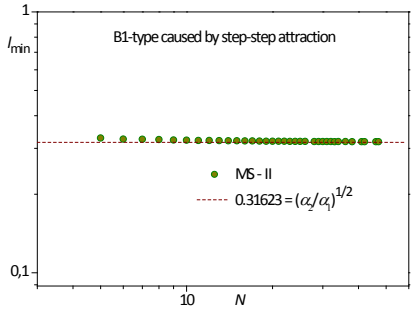 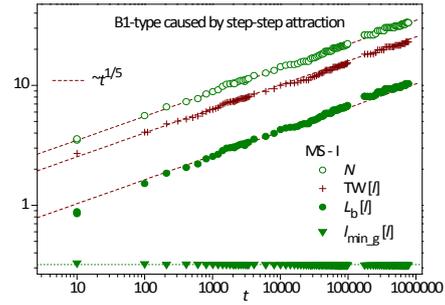

Figure 3. TE-model, MS-II, the minimal step-step distance in the bunch, measured in units of initial vicinal distance $l$, is not function of the number of steps in the bunch $N$ and thus the model is attributed to the B1-type.

Figure 4. TE-model, MS-I, the number of steps in the bunch $N$, the terrace width TW and the bunch width $L_b$ share the same time-scaling exponent, hence only one length-scale is needed what is typical for the B1-type. Data from several runs are used.

*B2-type.* – Some typical examples for experimental systems that show B2-type of step bunching, besides the most studied one – that of evaporating Si(111)-vicinals [1], are the KDP crystal growth [22, 23] and SiC epitaxial growth [24] although the latter shows quite rich bunching behavior and attribution to the B1-type is also possible after careful analysis of the experimental data using our systematic approach. I will illustrate this type with results on the model of diffusion limited vicinal evaporation already studied by Stoyanov and Tonchev [3] (called here EvEm-model) and the reason for this revisit is that the original study was restricted by the computational power at that time to the method of *single* bunch. Later on specially designed experiments [25] have shown that the process of Si(111) vicinal evaporation is actually controlled by the diffusion of the Si-adatoms on the terraces rather than by the attachment/detachment kinetics at the steps as suggested by a later study [16]. Very recently sophisticated experimental setup was developed [26] that would permit to study the time dependencies describing thoroughly the bunching process. As usually, systematic numerical studies play important role in planning experimental strategy and understanding its results.

The equations for step velocity are rather complex [3] and I will not adduce them here. Also, the original values of the parameters are preserved and only the parameter that contains the magnitude of the step-step repulsion is increased approximately thrice in order to permit faster computations. The reason is that in the method of *single bunch* used [3] one deals with systems of maximum 60-70 equations (steps) while in the present computation are included 1000 steps. The initial vicinal geometry comprises steps randomly deviated from their equidistant positions in order to permit development of the instability in a way similar to the real one. As for the previous, TE-model, these calculations include only the value $n=2$ from the step-step repulsion law, i.e. the repulsion energy decays with the inverse square of the step-step distance. The bunching process in the EvEm model is illustrated qualitatively by the step trajectories, Figure 5, and surface slope, Figure 6, and quantities that are monitored are marked.

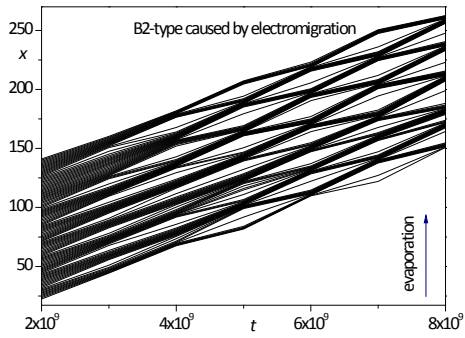
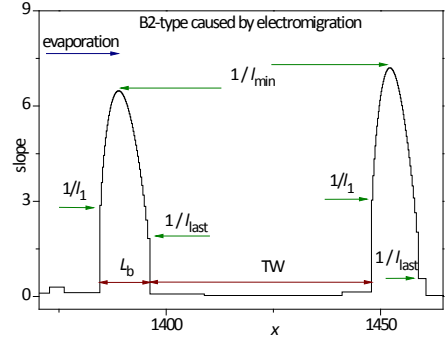

Figure 5. Step trajectories in the model of diffusion limited vicinal evaporation affected by electromigration of the adatoms (EvEm-model). Every step moves in the up direction mediating the vicinal surface evaporation. It is well seen how bunches split in two.

Figure 6. EvEm-model, surface profile for well-developed instability. With arrows are shown the places where appear some of the quantities used by the monitoring schemes. The higher slope is in the bunch with more steps (the right one), hence B2-type.

Next are shown the studies of the size-scaling (MS-II), first of the bunch width $L_b$, Figure 7, and then of the minimal, first and last step-step distance, Figure 8. Together with the studies of the time-scaling (MS-I) of the relevant quantities, Figure 9, these scaling exponents form a set that could be compared with the predictions of the universality classes in bunching hypothesis [16,27]. Here I will only stress that the comparison leads to the following conclusion: the set of *obtained* exponents with n=2, corresponds to the originally *predicted* for n=0 and thus is systematically shifted down in *n* with 2. There is still no explanation for such a shift and thus the present results are a challenge for further modification of the continuum equation which reflects the hypothesis of universality.

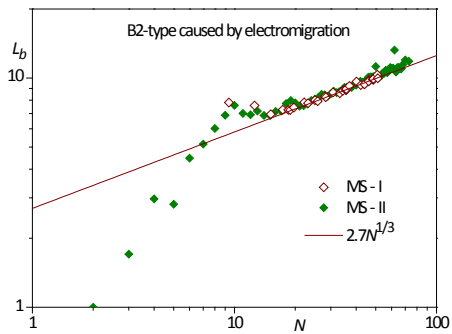
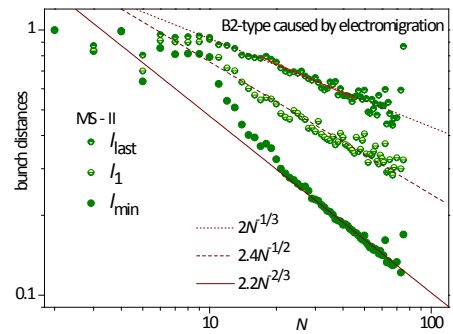

Figure 7. EvEm-model, bunch width $L_b$ versus number of steps *N* from both monitoring schemes. The slope of the guiding-eye line corresponds to size-scaling exponent of 1/3 (note the log-log character of the plot).

Figure 8. EvEm-model, MS-II, bunch distances - minimal, first and last, versus number of steps *N* as obtained from MS-II. The slopes of the guiding-eye lines correspond to size-scaling exponents of -2/3, -1/2 and -1/3 correspondingly. Note that also the last bunch distance decreases when *N* increases.

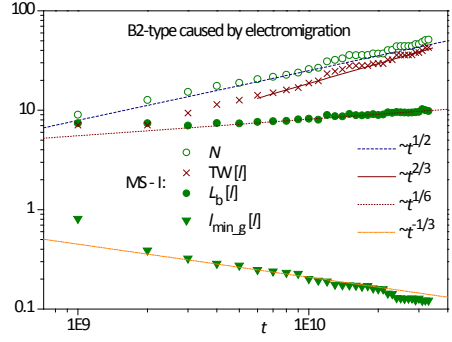

Figure 9. EvEm-model, MS-I, time dependencies of the number of steps *N*, terrace width TW, bunch width $L_b$, and the globally minimal step-step distance $l_{min\_g}$. The slopes of guiding-eye lines correspond to time-scaling exponents of 1/2, 2/3, 1/6 and -1/3, respectively.

Further comment on the results in this subsection is needed concerning the time-scaling exponent of the globally minimum step-step distance which is -1/3 as shown on Figure 9. The same exponent could be obtained if plugging in $t^{1/2}$ instead of *N* (time-scaling of *N*, shown on Figure 6) in the size-scaling relation of $l_{min}$ (as found on Figure 8):

$$l_{min} \sim N^{-2/3} \sim \left(t^{1/2}\right)^{-2/3} \sim t^{-1/3}$$

Thus one could argue that only the resulting scaling is enough to describe the spatiotemporal evolution of the vicinal surface but I will show in a subsequent study that the scaling above is invariant across two different B2-type models while the other scaling relations – in particular the size-scaling of $l_{min}$ and the time-scaling of *N* still distinguish unambiguously between the models.

**Perspective.** The nearest future of our computational studies of step bunching phenomena is to find the exact time-scaling of the number of steps in the bunch *N*, including the pre-factor, for the three models of B1-type available, all these having mathematically identical stabilizing and destabilizing terms in the equations for step velocity. As a preliminary study shows [21] namely the time-scaling of *N* could distinguish between the models and thus serve as a reference frame to plan, carry and understand experiments. In progress are also studies of three models of B2-type: diffusion and attachment-detachment limited evaporation affected by electromigration of the adatoms and a minimal model called MM0 which has the same destabilizing part in the equation(s) for step motion as in the latter but the stabilizing part is restricted to depend only on the widths of the two adjacent to the step terraces. Third direction of our studies comprises models that would eventually lead to what we now call B2m-type - simultaneous bunching *and* meandering (the steps are no more straight but wavy or *meandered*), which phenomenon is still anti-paradigmatic in a sense that in the paradigm the preconditions for bunching - normal Ehrlich-Schwoebel effect in (vicinal) evaporation *or* inverse Ehrlich-Schwoebel effect in (vicinal) growth, and meandering - normal Ehrlich-Schwoebel effect in growth *or* inverse Ehrlich-Schwoebel effect in evaporation cannot be realized simultaneously (downstep electromigration of the adatoms destabilizes both growing and evaporating vicinal surfaces). Nevertheless, there is sufficient experimental evidence both on metal [28,29] and semiconductor [30] vicinal surfaces for simultaneous bunching *and* meandering although the instability scenario is somewhat different. The bunches of B2m-type are expected to have their minimal step-step distance (largest slope) in the beginning of the bunches [14] rather than in the middle as seen on Figure 6 for the B2-type. Quite recently similar type of behavior was reported [31] in a KMC study of evaporating vicinal surface of GaN(0001).

**Acknowledgments.** Our present HPC facility MADARA was built with the prevailing financial support from the Bulgarian National Science Fund through grant No. ДОО02-52/RNF01/0110. Some financial support from the same fund through IRC CoSiM grant is also appreciated.

# Класификация на явленията на групиране

Веселин Тончев

*Институт по Физикохимия, Българска Академия на Науките*

*1113 София, България*

**Абстракт.** Класификацията на групирането на прави стъпала върху вицинални кристални повърхности обособява два типа според поведението на минималното разстояние в групата $l_{min}$ с нарастване на броя стъпала в нея $N$. В B1-типа $l_{min}$ остава постоянно докато в B2-типа то намалява. Двата типа са илюстрирани с нови резултати за известни модели. Точният числен анализ е насочен към режима на междинна асимптотика, в който се формират самоподобни пространствено-времеви структури. В модела на Терсоф и сътр. поредицата равноотдалечени стъпала е дестабилизирана от междустъпално привличане с безкраен обхват. Показано е, че този модел принадлежи към B1-типа и степента 1/5 е получена за времевото поведение на $N$, ширината на терасата между групите и ширината на самите групи. Една разширена поредица от степени е получена за модела на С. Стоянов на дифузионно-контролирано изпарение, повлияно от електромиграция на адатомите. Този модел е от B2-тип и показва систематично отместване на степените спрямо предсказанията на хипотезата за класове на универсалност в групирането, което налага по-нататъшна ѝ модификация.